\documentclass[12pt]{article}

\usepackage{cite}

\makeatletter
\def\asymp#1%
{\mathrel{\raisebox{-.4em}{$\widetilde{\scriptstyle #1}$}}}
\makeatother

\newcommand\nn         {\nonumber}

\newcommand\gs         {\ensuremath{g_{\mathrm{s}}}}
\newcommand\msbar      {\ensuremath{{\overline {\rm MS}}}}

\renewcommand\d        {{\mathrm d}}
\renewcommand\i        {{\mathrm i}}

\newcommand\Oe[1]      {\ensuremath{\mathrm O(\ep^{#1})}}

\newcommand{\cV}       {{\cal V}}
\def\ltap{\raisebox{-.4ex}{\rlap{$\,\sim\,$}} \raisebox{.4ex}{$\,<\,$}}
\def\gtap{\raisebox{-.4ex}{\rlap{$\,\sim\,$}} \raisebox{.4ex}{$\,>\,$}}

\def\ep{\epsilon}

\def\beq{\begin{equation}}
\def\eeq{\end{equation}}
\def\beeq{\begin{eqnarray}}
\def\eeeq{\end{eqnarray}}
\def\cm{{\cal A}}

\def\average#1{#1}
\def\tq{{\tilde q}}
\def\tg{{\tilde g}}

\def\bom#1{{\mbox{\boldmath $#1$}}}
\def\to{\rightarrow}

\def\RS{{\scriptscriptstyle\rm R\!.S\!.}}

\def\gtilde{{\tilde \gamma}}

\def\CDR{{\rm CDR}}

\def\DR{{\rm DR}}
\newcommand{\la}{\langle}
\newcommand{\ra}{\rangle}

\def\nn{\nonumber}

\def\AP{Altarelli--Parisi }
\def\ID{1 \kern -.45 em 1}
\def\nquad{\!\!\!\!\!\!\!}

\def\draftdate{\relax}
\def\mda{\relax}
\def\mua{\relax}
\def\mla{\relax}
\def\draft{
\def\thtystars{******************************}
\def\sixtystars{\thtystars\thtystars}
\typeout{}
\typeout{\sixtystars**}
\typeout{* Draft mode!
         For final version remove \protect\draft\space in source file *}
\typeout{\sixtystars**}
\typeout{}
\def\draftdate{\today}
\def\mua{\marginpar[\boldmath\hfil$\uparrow$]%
                   {\boldmath$\uparrow$\hfil}%
                    \typeout{marginpar: $\uparrow$}\ignorespaces}
\def\mda{\marginpar[\boldmath\hfil$\downarrow$]%
                   {\boldmath$\downarrow$\hfil}%
                    \typeout{marginpar: $\downarrow$}\ignorespaces}
\def\mla{\marginpar[\boldmath\hfil$\rightarrow$]%
                   {\boldmath$\leftarrow $\hfil}%
                    \typeout{marginpar: $\leftrightarrow$}\ignorespaces}
\overfullrule 5pt
\oddsidemargin -15mm
\marginparwidth 29mm
}

\def\stars{\strut\leaders\hbox{*}\hfill\strut}
\def\starline{\hfil\strut\hfil\hbox to \textwidth {\stars}\hfil}

\begin{document}

\begin{titlepage}
\renewcommand{\thefootnote}{\fnsymbol{footnote}}
\begin{flushright}
CERN--TH/2000-336\\ 
BI-TP 2000/29 \\  hep-ph/0011222 
     \end{flushright}
\par \vspace{10mm}
\begin{center}
{\Large \bf
One-loop Singular Behaviour of QCD and\\[1ex]
SUSY QCD Amplitudes with Massive Partons} 
\end{center}
\par \vspace{2mm}
\begin{center}
{\bf Stefano Catani}~$^1$\footnote{On leave of absence from INFN,
Sezione di Firenze, Florence, Italy.}{\bf ,}
{\bf Stefan Dittmaier}~$^2$ and
{\bf Zolt\'an Tr\'ocs\'anyi}~$^3$\footnote{Sz\'echenyi fellow of the
Hungarian Ministry of Education.}

\vspace{5mm}

{$^1$ Theory Division, CERN, CH-1211 Geneva 23, Switzerland} \\[.5em]
{$^2$ Theoretische Physik, Universit\"at Bielefeld, 
D-33615 Bielefeld, Germany} \\[.5em]
{$^3$ University of Debrecen and \\Institute of Nuclear Research of
the Hungarian Academy of Sciences\\ H-4001 Debrecen, PO Box 51, Hungary}

\vspace{5mm}

\end{center}

\par \vspace{2mm}
\begin{center} {\large \bf Abstract} \end{center}
\begin{quote}
\pretolerance 10000

We discuss the structure of infrared and ultraviolet singularities in
on-shell QCD and supersymmetric QCD amplitudes at one-loop order.
Previous results, valid for massless partons, are extended to the case
of massive partons. Using dimensional regularization, we present a
general factorization formula that controls both the singular
$\ep$-poles and the logarithmic contributions that become singular for
vanishing masses.  We introduce generalized Altarelli--Parisi splitting
functions and discuss their relations with the singular terms in the
amplitudes. The dependence on the regularization scheme is also considered.
 
\end{quote}

\vspace*{\fill}
\begin{flushleft}
     CERN--TH/2000-336 \\   November 2000
\end{flushleft}
\end{titlepage}

\renewcommand{\thefootnote}{\fnsymbol{footnote}}

\section{Introduction}
\label{intro}

Hard-scattering processes that lead to multiparton final states
with heavy particles are important for present physics studies within and 
beyond the Standard Model. Their importance will further increase at future
high-energy colliders (see, e.g., Refs.~\cite{acc, Altarelli:2000ye} and references
therein). Reliable theoretical predictions for these processes require
the evaluation of (at least) next-to-leading order QCD corrections,
independently of the nature (QCD, electroweak, SUSY) of the interaction
that produces these processes at the lowest perturbative order. The computation
of QCD radiative corrections to cross sections that involve heavy partons
are certainly more complicated than in the case of massless partons,
as known since the first NLO calculations
of heavy-quark production in hadron collisions
\cite{HQNLO}.
 
Most of the available techniques (see, e.g., the list of references in
Sect.~4 of the QCD Chapter of Ref.~\cite{Altarelli:2000ye}) to perform
NLO calculations require the analytic evaluation of the infrared (soft
and collinear) and ultraviolet singularities of the one-loop amplitudes
for the corresponding process.  The knowledge of the singularity
structure in a general (process-independent) form is thus useful for
several reasons.  Loop calculations are always very cumbersome and
error-prone, therefore, knowing any general property of the result is a
good check on the calculations. The knowledge of the singularities of
the loop amplitude can also be used as a possible tool to split the
loop calculation into
a divergent part, which is known in analytic form, and  a finite part,
to be evaluated by numerical methods.  In the case of infrared-safe
observables the infrared singularities of the real and loop corrections
must cancel, therefore, knowing the singularity structure without
performing the explicit loop computation helps devising
process-independent calculational schemes at NLO.

The singular behaviour of QCD amplitudes involving massless partons is
completely known at one-loop order \cite{GG,KSTsing,CSdipole}, and the
coefficients of the divergent
$1/\ep^n$ poles for $n=$2, 3 and 4 are also known for two-loop amplitudes
in the most general case \cite{C2loop}. Despite the numerous one-loop
calculations involving massive particles that have been performed so far,
to our knowledge the general form of the singularities in 
amplitudes with massive partons has not yet been presented 
even at one-loop order.

In this paper we discuss the structure of the singularities of one-loop
QCD (or QED) corrections to on-shell scattering amplitudes
that involve both massless and massive partons. By singularities
we mean not only the terms that diverge in absence of regularization, but also 
those that become divergent in the limit of vanishing parton masses. 
With increasing collision energies the ratios of the particle masses to
the kinematic (Mandelstam) invariants tend to zero. Thus the amplitude contains
logarithms of such ratios that become large and whose explicit control is
important for numerical stability. 
This feature has particular relevance for electroweak physics and 
one-loop QED corrections where collinear singularities
are physically regularized by the small values of the light-fermion masses. 

We present a universal (process-independent) factorization formula that
embodies the structure and the coefficients of all these singular terms
in the one-loop amplitudes.  We use dimensional regularization in
$d=4-2\ep$ dimensions for both ultraviolet and infrared divergences,
and we thus give the coefficients\footnote{These coefficients can also
be obtained by using the general results of Ref.~\cite{Keller:1999tf}.}
of the corresponding poles $1/\ep^2$ and $1/\ep$.  The factorization
formula also explicitly exhibits all logarithmically-enhanced terms
$\ln^2 m$ and $\ln m$ that become singular for vanishing mass $m$ of
the massive partons. Furthermore, we can have full control of all the
terms that are not analytic in the massless limit, because the
factorization formula also contains the constant (when $m \to 0$) terms
that originate from the non-commutativity of the limits $m \to 0$ and
$\ep \to 0$. Therefore, in the limit $m \to 0$, the finite (when
$\ep \to 0$) one-loop contribution that is not included in the
factorization formula tends smoothly to the finite contribution of the
corresponding massless amplitude.  

There are two ways to derive the singular terms.  They can be obtained
either by explicitly performing loop calculations or by using unitarity.
We use the latter method. It amounts to exploiting the fact that the
infrared singularities of the real and loop corrections cancel each
other in the computation of infrared-safe observables. Thus, the
singular structure of the loop corrections can be derived from that of
the real corrections.  Our process-independent calculation of the real
corrections is based on the dipole subtraction formalism\cite{CSdipole}.
In particular, we have explicitly extended the results of
Refs.~\cite{CSdipole, di00} to the general case of partons with
arbitrary masses.  In this paper we present only the final results on
the singular behavior of one-loop amplitudes.  Complete details of the
extended formalism will be presented elsewhere.

After fixing our notation and conventions 
in Sect.~\ref{notat}, we present our 
results on the singular terms of QCD and SUSY QCD amplitudes in
Sect.~\ref{sing1loopmassive}. 
The singularities 
produced by QED corrections to electroweak and QCD amplitudes can be
obtained from the QCD results as a special case.
In Sect.~\ref{APrelation} we relate the singularities to individual
terms in generalized \AP functions. Section~\ref{sum} contains our
summary, and the Appendix provides a list of generalized \AP functions.

\section{Notation}
\label{notat}

\subsection{Dimensional regularization}
\label{dimreg}

In the evaluation of loop amplitudes one encounters ultraviolet and
infrared divergences that have to be properly
regularized. The most efficient method to simultaneously regularize
both types of singularities in gauge theories is to use dimensional
regularization \cite{tHV}.
The key ingredient of
dimensional regularization is the analytic continuation of loop momenta
to $d=4-2\ep$ space-time dimensions.  Having done this, one is left
with some freedom regarding the dimensionality of the momenta of the
external particles as well as the number of polarizations of both
external and internal particles. This leads to different regularization
schemes (RS) within the dimensional-regularization prescription
(see, e.g., Refs.~\cite{KST2to2, CSTrsdep}).

The two variants of dimensional regularization that are mostly used in
one-loop computations are conventional dimensional regularization (CDR)
and dimensional reduction (DR). In both schemes one considers 2
helicity states for spin-$\frac{1}{2}$ Dirac fermions.  The essential
difference between the two schemes regards the number of the helicity
states of the gluons in the loop.  The gluon has $d-2$ helicities in
CDR and 2 helicities in DR. Since the number of helicity states of
gluons and Dirac fermions is the same in DR,
this scheme preserves supersymmetric Ward identities.  Within each
scheme one can still choose the external particles (their momenta and
helicities) in the amplitudes to be either $d$-dimensional or 4-dimensional. 
At one-loop order, these choices lead to differences of ${\cal O}(\ep)$,
which do not have any effect on the results presented in this paper.

Note that ultraviolet and infrared divergences 
behave differently with respect to
di\-men\-sional-re\-gu\-la\-ri\-za\-tion prescriptions. The ultraviolet RS
dependence can ultimately be removed by a proper redefinition of
the renormalized running coupling. The infrared RS dependence instead leads
to contributions (which are not vanishing for $\ep \to 0$) that depend
on the specific amplitude and that cannot be reabsorbed by an overall
(i.e., independent of the amplitude) redefinition of the renormalized
coupling. These features, which were explicitly 
pointed out in Ref.~\cite{KST2to2},
will be discussed in detail in the following sections.
Note also that, in the calculation of physical quantities, the RS dependence of
loop amplitudes has to be consistently matched to that of tree amplitudes.
This issue is discussed on quite a general basis in
Ref.~\cite{CSTrsdep}.

\subsection{Partially renormalized amplitudes}
\label{ren}

We consider amplitudes $\cm_m$ that involve $m$ external coloured
particles (gluons, massless and massive quarks, gluinos and squarks)
with momenta $p_1, \dots, p_m$, masses $m_1, \dots, m_m$ and an
arbitrary number and type of colourless particles (photons, leptons,
vector bosons, etc.).  We always consider the amplitudes in the
crossing-symmetric, but unphysical channel when all particle momenta
are outgoing. The amplitudes are denoted by $\cm_m(p_1,m_1,\dots,p_m,m_m)$
(or, shortly, $\cm_m(\{p_i, m_i\})$), and the dependence on the momenta
and quantum numbers of the colourless particles is always understood.

For amplitudes of processes involving massive particles the \msbar\
subtraction scheme is not always used to perform charge (coupling)
renormalization (see, e.g., Ref.~\cite{chargeren}). To
leave the freedom of choosing a favourite charge-renormalization
scheme, we find it convenient to use {\em mass-renormalized}, but
{\em charge-unrenormalized amplitudes}. Thus, in the perturbative
expansion
\beq
\label{loopex}
\cm_m(\gs, \mu^2; \{p_i,m_i\}) =
\left(\frac{\gs\mu^\ep}{4\pi} \right)^q
\,\left[\,\cm_m^{(0)}(\{p_i,m_i\}) 
+ \left(\frac{\gs}{4\pi}\right)^2\,\cm_m^{(1)}(\mu^2; \{p_i,m_i\})
+ {\rm O}(\gs^4)\right]
\eeq
\gs\ stands for the {\em bare} strong coupling and $m_i$ are the
{\em renormalized} mass parameters 
(i.e., masses and related parameters such as
those which appear in Yukawa couplings). The renormalized masses are obtained
from the bare masses $m_i^{(0)}$ by the replacement $m_i^{(0)} \to m_i =
m_i^{(0)} + \gs \,\delta m_i$, so that the (ultraviolet-divergent)
mass renormalization constants $\delta m_i$ are implicitly contained 
in $\cm_m^{(1)}$.
In Eq.~(\ref{loopex}) $q$ is a non-negative integer, and $\mu$ is the 
dimensional-regularization scale. 

Equation~(\ref{loopex}) fixes the normalization of the tree-level, $\cm_m^{(0)}$,
and one-loop, $\cm_m^{(1)}$, coefficient amplitudes that we use in the
rest of the paper\footnote{Precisely speaking, $\cm_m^{(0)}$ is not
necessarily a tree amplitude, but rather the lowest-order amplitude for
a given process; $\cm_m^{(1)}$ is the corresponding NLO 
correction. For instance, in the case of $gg \to \gamma \gamma$,
$\cm^{(0)}$ involves a quark loop.}.
Although it is not explicitly denoted in Eq.~(\ref{loopex}),  
$\cm^{(0)}$ and $\cm^{(1)}$ are both dependent on the RS.

\subsection{Colour space}
\label{col}

We shall present the singular structure of the QCD 
and SUSY QCD amplitudes directly in
colour space. In particular, we use the same notation as in
Ref.~\cite{CSdipole}.

The colour indices of the $m$ partons in the amplitude $\cm_m$ are
generically denoted by $c_1,\dots,c_m$: $c_i=\{a \}= 1,\dots,N_c^2-1$
for particles in the adjoint representation (gluons, gluinos) and
$c_i=\{\alpha\}=1,\dots,N_c$ for particles in the fundamental
representation (quarks, squarks and their antiparticles) of the gauge
group.  We formally introduce an orthogonal basis of unit vectors $\{
|c_1,\dots,c_m \ra \}$ in the $m$-parton colour space, in such a way
that the colour amplitude can be written as follows:
\beq
\label{medef}
\cm_m^{c_1,\dots,c_m}(p_1,m_1,\dots,p_m,m_m) \equiv
\la c_1,\dots,c_m \, | \, \cm_m(p_1,m_1,\ldots,p_m,m_m)\ra \:.
\eeq
Thus $|\cm_m(p_1,m_1,\ldots,p_m,m_m)\ra$ is an abstract vector in
colour space, and the square amplitude summed over colours is
\beq
|\cm_m(\{p_i,m_i\})|^2 = 
\la \cm_m(\{p_i,m_i\}) \, | \, \cm_m(\{p_i,m_i\}) \ra \:.
\eeq

Colour interactions at the QCD vertices are represented by associating
a colour charge ${\bf T}_i$ with the emission of a gluon from
each parton $i$.  The colour charge ${\bf T}_i= \{T_i^a \} $ is a vector
with respect to the colour indices $a$ of the emitted gluon and
an $SU(N_c)$ matrix with respect to the colour indices of the parton $i$.
More precisely, its action onto the colour space is defined by
\beq
\la c_1,\dots,c_i,\dots,c_m \,| \, T_i^a \, | \,b_1,\dots,b_i,\dots,b_m \ra 
= \delta_{c_1b_1} \dots T_{c_ib_i}^a \dots \delta_{c_mb_m} \;,
\eeq
where $T_{c b}^a$ is the colour-charge matrix in the representation of
the final-state emitting particle $i$, i.e.\ $T_{c b}^a=i f_{cab}$ if $i$ is a
gluon or a gluino, $T_{\alpha\beta}^a=t_{\alpha\beta}^a$ if $i$ is a
(s)quark, and $T_{\alpha\beta}^a=-t_{\beta\alpha}^a$ if $i$ is an
anti(s)quark. The colour-charge operator of an initial-state parton is
defined by crossing symmetry,
that is by $({\bf T}_i)^a_{\alpha \beta} 
= - t^a_{\beta \alpha }$ if $i$ is an initial-state 
(s)quark  
and $({\bf T}_i)^a_{\alpha \beta} =
t^a_{\alpha \beta}$ if $i$ is an initial-state
anti(s)quark.

In this notation, each vector $|\cm_m(p_1,m_1,\dots,p_m,m_m) \ra$ is a
colour singlet, so colour conservation is simply
\beq
\label{colcon}
\sum_{i=1}^m {\bom T}_i \;|\cm_m \ra = 0 \;.
\eeq

The colour-charge algebra for the product 
$({\bf T}_i)^a ({\bf T}_j)^a \equiv {\bf T}_i \cdot {\bf T}_j$ is:
\beq
{\bom T}_i \cdot {\bom T}_j ={\bom T}_j \cdot {\bom T}_i \;\;\;\;{\rm if}
\;\;i \neq j; \;\;\;\;\;\;{\bom T}_i^2= C_i \;,
\eeq
where $C_i$ is the quadratic Casimir operator in the representation of
particle $i$, and we have $C_F= T_R(N_c^2-1)/N_c= (N_c^2-1)/(2N_c)$ in the
fundamental and $C_A=2\,T_R N_c=N_c$ in the adjoint representation,
i.e.~we are using the customary normalization $T_R=1/2$.

We remind the reader that, in the cases of amplitudes with $m=2$ and
$m=3$ coloured partons, the colour-charge algebra can always be recast
in a fully factorized form in terms of Casimir operators of the $m$
partons (see the Appendix A of the second paper in
Ref.~\cite{CSdipole}).
 
Note that in the following sections we always refer to one-loop QCD
corrections to scattering amplitudes.  Nonetheless, most of the results
can straighforwardly be used for the case of one-loop QED corrections.
To this purpose it is sufficient to replace the colour couplings
$\gs {\bom T}_i$ by the electric couplings $g e_i$, where $g$ is the
electric charge of the electron and $e_i$ is the charge of the parton
$i$ in units of the electron charge.  In terms of colour factors, this
implies the replacements $C_F \to e_i^2$, $T_R \to 1$ and $C_A \to 0$.

\section{Singular behaviour at one-loop order with massive partons}
\label{sing1loopmassive}

\subsection{QCD amplitudes}
\label{sing1loopQCD}

In this subsection we present our results on the singular
behaviour of QCD amplitudes at one-loop order. In the massless case the
one-loop coefficient subamplitude $\cm_m^{(1)}$ has double and single
poles in $\ep$ that can be obtained by a process-independent factorization
formula \cite{GG, KSTsing, CSdipole}. Similar poles, although with different
coefficients, appear if the amplitude involves massive partons. We find
that these singularities are still universal, so that the factorization formula
for the massless amplitudes can be generalized to the massive case.

The general factorization formula for the one-loop coefficient
subamplitude $\cm_m^{(1)}$ is%
\footnote{Here and in the following, the labels ${\rm R.S.}$ explicitly
denote the RS~dependence of the various quantities.}
\beq
\label{ff1loop}
| \cm_m^{(1)}(\mu^2;\{p_i,m_i\}) \ra_{\RS} =
{\bom I}_m^{\RS}(\ep,\mu^2;\{p_i,m_i\})\,
| \cm_m^{(0)}(\{p_i,m_i\}) \ra_{\RS}
+ | \cm_m^{(1)\, {\rm fin}}(\mu^2;\{p_i,m_i\}) \ra
+\Oe{}\:.
\eeq
All the $\ep$-poles are included in the factor ${\bom I}$, so the remaining
contributions on the right-hand side can be safely expanded in $\ep$ for 
$\ep \to 0$. Moreover, the factor ${\bom I}$ includes also the constant
(when $\ep \to 0$) terms related to the RS dependence. Thus, the
contribution $\cm_m^{(1)\, {\rm fin}}$ is not only finite, but it is also
RS-independent.

These features are shared by massless and massive amplitudes. In the
case of massive quarks, moreover, our factorization formula has
additional important properties related to logarithmically-enhanced
contributions. In the limit of one or more vanishing masses $m_i$, the
subamplitude $\cm_m^{(1)}(\mu^2; \{p_i,m_i\})$ contains {\em logarithmic}
terms of the type $\ln^2 m_i$ and $\ln m_i$ that become singular. It
also contains {\em constant} terms that originate from the
non-commutativity of the limits $m_i \to 0$ and $\ep \to 0$. We are
able to embody all these logarithmic and constant terms\footnote{In the
small-mass limit, the mass $m_i$ replaces dimensional regularization as
regulator of collinear singularities. In this sense, we can say that we
can control the ensuing regularization-scheme dependence including
finite (when $m_i \to 0$) terms.} in the factor ${\bom I}$. Thus, in
the limit $m_i \to 0$  the finite contribution 
$\cm_m^{(1)\, {\rm fin}}(\dots, p_i,m_i, \dots)$ tends smoothly to the
finite contribution $\cm_m^{(1)\, {\rm fin}}(\dots, p_i,m_i=0, \dots)$
of the corresponding amplitude in the theory where the parton $i$ is massless:
\beq
\label{masslesslimit}
\lim_{m_i \to 0} 
\cm_m^{(1)\, {\rm fin}}(\mu^2; p_1,m_1, \dots, p_i,m_i, \dots)  
= \cm_m^{(1)\, {\rm fin}}(\mu^2; p_1,m_1, \dots, p_i,m_i=0, \dots)  \:.
\eeq
Note that Eq.~(\ref{masslesslimit}) is valid independent of the actual
definition of the renormalized mass $m_i$ (or of the related Yukawa
couplings).  We can use either the pole-mass definition or the \msbar\ 
definition, because the terms $\ln m_i$ originating from the different
definitions are always suppressed by the corresponding mass factor as
$m_i \ln m_i$.

The factorization formula (\ref{ff1loop}) and the property in
Eq.~(\ref{masslesslimit}) can be used to check the calculation of the
massive amplitude by comparing it to the corresponding massless
calculation. In the asymptotic regime where $m_i$ is much smaller than
any of the relevant kinematic invariants $Q$, these equations can also
be used to directly obtain (apart from corrections of ${\cal O}(m_i/Q)$)
the one-loop massive amplitude from the corresponding massless
amplitude, without explicitly computing the former.

The first term on the right-hand side of Eq.~(\ref{ff1loop}) has a
factorized structure in colour space. The singular dependence (poles in
$\ep$ and logarithms in $m_i$) is embodied in the factor 
${\bom I}_m^\RS$ that acts as a colour-charge operator onto the colour
vector $| \cm_m^{(0)} \ra_\RS$. Note that both factors are
RS-dependent.  In particular, the product of the RS-dependent terms
of O$(\ep)$ in $\cm_m^{(0)}$ and double poles $1/\ep^2$ in ${\bom I}$
produces, in general, an RS dependence of $\cm_m^{(1)}$ that begins at
O$(1/\ep)$.

The explicit expression for ${\bom I}_m$ in terms of the colour charges
of the $m$ partons is the following:
\beeq
\label{iee}
&&\nquad
{\bom I}_m^\RS(\ep,\mu^2;\{p_i,m_i\}) =
\frac{(4\pi)^\ep}{\Gamma(1-\ep)} \Bigg\{
q\,\frac12\left( \frac{\beta_0}{\ep} - {\tilde{\beta}}_0^{\RS} \right)
\\ \nn &&\qquad\qquad
+ \sum_{j,k=1 \atop k \neq j}^m {\bom T}_j\cdot{\bom T}_k
\left(\frac{\mu^2}{|s_{jk}|} \right)^{\ep} \left[
\cV^{(\rm cc)}_{jk}(s_{jk};m_j,m_k;\ep)
+ \frac{1}{v_{jk}}\,
\left(\frac{1}{\ep}\,\i\pi\,
- \frac{\pi^2}{2}\right) \Theta(s_{jk}) \right]
\\ \nn &&\qquad\qquad
- \sum_{j=1}^m \Gamma_j^{\RS}(\mu,m_j;\ep)
 \Bigg\}\:.
\eeeq

The first term on the right-hand side of Eq.~(\ref{iee}) contains the
ultraviolet divergences, to be removed by the renormalization of the
bare coupling $\gs$.  It is proportional to $q$, which is the overall
power of $\gs$ in Eq.~(\ref{loopex}), and $\beta_0$ is the first
coefficient of the QCD beta function
\beq
\label{beta0}
\beta_0 =
\frac{11}{3}\,C_A - \frac{4}{3}\,T_R ( N_f + N_F)\:,
\eeq
where $N_f$ and $N_F$ are the numbers of {\em massless} and
{\em massive} quark flavours, respectively.  The constant coefficient
${\tilde{\beta}}_0^{\RS}$ parametrizes the ultraviolet RS~dependence.
Setting ${\tilde{\beta}}_0^{\CDR}=0$ by definition in the CDR scheme,
its corresponding value in the DR scheme is\cite{ACMP}
(see also Ref.~\cite{KST2to2})
\beq
\label{betat}
{\tilde{\beta}}_0^{\DR}= \frac{1}{3} \,C_A \;.
\eeq

The second and third terms on the right-hand side of Eq.~(\ref{iee})
have an infrared origin.  We have determined them by exploiting the
fact that, in NLO calculations of infrared-safe cross sections, the
infrared singularities of the one-loop amplitudes are cancelled by
their counterpart in the real-emission contribution. The latter has
been computed by extending the dipole subtraction formalism
\cite{CSdipole, di00} to the case of massive partons.  The infrared
contribution that leads to colour correlations proportional to
${\bom T}_j\cdot{\bom T}_k$ is produced by soft {\em and} collinear
singularities. The infrared contributions $\Gamma_j^{\RS}$ are produced
by either collinear (but not soft) or soft (but not collinear) singularities.

The singular function $\cV^{(\rm cc)}_{jk}$ that controls colour
correlations is symmetric with respect to $j \leftrightarrow k$ and
depends on the Lorentz invariant $s_{jk} = 2p_j\cdot p_k$ and on the
parton masses. In particular, it depends on the the relative
velocity $v_{jk}$ of particles $j$ and $k$:
\beq
\label{relvelkq}
v_{jk} = \sqrt {1 - \frac{m_j^2 m_k^2}{(p_jp_k)^2}}\:.
\eeq
Its explicit
expression for non-vanishing masses $m_j$ and $m_k$ is
\beq
\label{cVmassive}
\cV^{(\rm cc)}_{jk}(s_{jk};m_j,m_k;\ep) =
\frac{1}{2\ep} \,\frac1{v_{jk}}\, \ln\frac{1 - v_{jk}}{1 + v_{jk}} 
- \frac{1}{4} 
\Bigg( \ln^2\!\frac{m_j^2}{|s_{jk}|} + \ln^2\!\frac{m_k^2}{|s_{jk}|} \Bigg)
-\frac{\pi^2}{6} \:,
\eeq
while for one or two vanishing masses we find
\beeq
&& 
\label{cVmasslessk}
\cV^{(\rm cc)}_{jk}(s_{jk};m_j,0;\ep) = \frac{1}{2\ep^2} +
\frac{1}{2\ep} \ln\frac{m_j^2}{|s_{jk}|}
- \frac{1}{4} \ln^2\!\frac{m_j^2}{|s_{jk}|}
-\frac{\pi^2}{12} \:,
\\ && 
\label{cVmasslessj}
\cV^{(\rm cc)}_{jk}(s_{jk};0,0;\ep) = \frac{1}{\ep^2} \:.
\eeeq

The functions $\Gamma_j^{\RS}$ depend on the flavour of the 
parton $j$ and on the parton masses. In the case of gluons and 
massless quarks (antiquarks) we have
\beeq
\label{cgammag}
\Gamma_g^{\RS}(\mu,m_{\{F\}};\ep) &=&
\frac{1}{\ep} \; \gamma_g - \gtilde_g^{\RS} -
\frac{2}{3}\,T_R \sum_{F=1}^{N_F} \;\ln\frac{m_F^2}{\mu^2}  \;,
\\
\label{cgammaq0}
\Gamma_q^{\RS}(\mu,0;\ep) &=& \frac{1}{\ep} \; \gamma_q - \gtilde_q^{\RS} \;,
\eeeq
while for massive quarks (antiquarks) we find
\beq
\label{cgammaqm}
\Gamma_q(\mu,m_q;\ep)= 
{\bom T}_q^2 \left( \frac{1}{\ep} - \ln\frac{m_q^2}{\mu^2} -2 \right) +
\gamma_q \,\ln\frac{m_q^2}{\mu^2} =
C_F \left[ \frac{1}{\ep} + 
\frac{1}{2} \ln\frac{m_q^2}{\mu^2} -2 \right] \;.
\eeq
The flavour coefficients $\gamma_j$ in Eqs.~(\ref{cgammag}) and 
(\ref{cgammaq0}) 
are
\beq
\label{cgamma}
\gamma_{j=q,\bar q} = 
\frac{3}{2}\,C_F\:,\qquad\qquad\quad\;\;
\gamma_g = \frac{11}{6}\,C_A - \frac{2}{3}\,T_R \, N_f \:.
\eeq
The coefficients $\gtilde_j^{\RS}$ parametrize the finite (for $\ep \to 0$)
contributions related to the RS~dependence of the one-loop amplitudes 
with external massless partons (the massive-quark function in 
Eq.~(\ref{cgammaqm}) does not depend on the actual version of dimensional
regularization used in the calculation).
The transition coefficients $\gtilde_j^{\RS}$ that
relate the RS mostly used in one-loop computations were first
calculated in Ref.~\cite{KST2to2}. They are given by 
\beq
\label{cgammatilde}
\tilde{\gamma}^{\rm CDR}_j = 0 \:, \qquad
\tilde{\gamma}^{\rm DR}_{j=q,{\bar q}} = \frac{1}{2} \,C_F \:, \qquad
\tilde{\gamma}^{\rm DR}_{j=g} = \frac{1}{6} \,C_A.
\eeq

Note that the dependence on the RS of the 
ultraviolet and infrared contributions is different. The ultraviolet
contribution ${\tilde{\beta}}_0^{\RS}$ in Eq.~(\ref{iee}) is proportional 
to the overall power $q$ of $\gs$ that controls the amplitude
(see Eq.~(\ref{loopex})) and it can ultimately be reabsorbed
by a process-independent redefinition of the renormalized coupling. 
The infrared contributions $\gtilde_j^{\RS}$ to Eq.~(\ref{iee}) instead
depend on the number and flavour of the external massless partons in the amplitude.
In the calculation of physical (infrared-finite) quantities this dependence
has to be cancelled by computing the corresponding real-emission contributions
in a consistent manner, that is, by using the same dimensional-regularization
prescription as in one-loop amplitudes \cite{CSTrsdep}.

Some comments on the structure of these results, in particular
about the $\ep$ poles and the mass logarithms, are appropriate.
\begin{itemize}
\item
If there are only massless partons, our result agrees with those in
Refs.~\cite{KSTsing,GG,CSdipole}. The structure of the $\ep$ poles
for massive quarks agrees with the results of Ref.~\cite{Keller:1999tf} and 
with the QED case considered in Ref.~\cite{di00}.
\item
The double poles $1/\ep^2$ in Eq.~(\ref{ff1loop}) are factorized
completely and not only in colour space. More precisely, there is a 
contribution $- {\bom T}_j^2/\ep^2$ from each external {\em massless} partons 
$j$. The simplest way to see that
is to expand Eq.~(\ref{iee}) in powers of $\ep$ and then use the colour
conservation relation (\ref{colcon}),
i.e.~$\sum_{k \ne j} {\bom T}_k = - {\bom T}_j$. One obtains the result
\beq
\label{ieeexp}
{\bom I}_m(\ep,\mu^2;\{p_i,m_i\}) = 
\sum_{j\atop m_j = 0} \frac{1}{\ep^2} \,
\sum_{k \ne j} {\bom T}_j \cdot {\bom T}_k
+ {\rm O}(1/\ep) = - \frac{1}{\ep^2} \sum_{j\atop m_j = 0} {\bom T}_j^2
+ {\rm O}(1/\ep) 
\eeq
that explicitly shows the absence of colour correlations at O$(1/\ep^2)$.
Nonetheless, single poles $1/\ep$ are both colour- and velocity-correlated.
\item
The term proportional to $1/\ep$ in Eq.~(\ref{cVmassive}) is familiar
from QED bremsstrahlung, and the imaginary part in Eq.~(\ref{iee}) is
the corresponding Coulomb phase.
\item
Since we have
\beq
\frac{1 - v_{jk}}{1 + v_{jk}} \; \asymp{m_j \to 0} \;
\frac{m_j^2m_k^2}{s_{jk}^2}\:,
\eeq
Eqs.~(\ref{cVmassive}), (\ref{cVmasslessk}) and (\ref{cVmasslessj})
are related by the following formal correspondence
\beq
\frac{1}{2\ep} \ln\frac{m_j^2}{|s_{jk}|}
- \frac{1}{4} \ln^2\!\frac{m_j^2}{|s_{jk}|}
-\frac{\pi^2}{12} + {\cal O}\left(\frac{m_j^2}{|s_{jk}|}\right)
\longleftrightarrow \frac{1}{2\ep^2}
\eeq
between mass logarithms in the massless limit and $1/\ep^2$ poles in
the massless theory. 
\item The various constant (for $\ep \to 0$) terms in
Eqs.~(\ref{cVmassive})--(\ref{cgammaqm}) are important to guarantee the
smooth limit in Eq.~(\ref{masslesslimit}).
One can always include additional finite terms in ${\bom I}$, provided
they are smooth in the massless limit. We chose to include the terms
that are proportional to $\pi^2 \Theta(s_{jk})$ in Eq.~(\ref{iee}).

\end{itemize}

We also add some other comments on the origin of the last two
terms on the right-hand side of Eq.~(\ref{iee}).  The contributions
$\Gamma_j^{\RS}$ can be related to the \AP\ splitting functions 
(see Sect.~\ref{APrelation}).  As already mentioned, we have evaluated
the colour-correlation term in Eq.~(\ref{iee}) by computing the
corresponding bremsstrahlung contribution. This directly gives the real
part of this term, namely the function $\cV^{(\rm cc)}_{jk}$.  To
obtain the corresponding imaginary part in the square bracket on the
right-hand side of Eq.~(\ref{iee}), we have exploited the fact that
the singular (real and imaginary) part of the colour-correlation term 
is directly proportional to the the following three-point
function:
\beq
\label{3pfun}
{\bom T}_j\cdot{\bom T}_k \int d^dq \;\frac{i}{q^2+i0} \;
\frac{p_j p_k}{[(p_j + q)^2 - m_j^2 +i0] \; [(p_k - q)^2 - m_k^2 +i0]} \;\;.
\eeq
The contribution of the gluon pole, 
\beq
\frac{i}{q^2+i0} \to 2\pi \delta_+(q^2) \;,
\eeq
gives $\cV^{(\rm cc)}_{jk}$, while the contribution from the poles in the 
massive propagators,
\beq 
\frac{1}{[(p_j + q)^2 - m_j^2 +i0] \; [(p_k - q)^2 - m_k^2 +i0]}
\to - 4\pi^2 \delta_+((p_j + q)^2 - m_j^2) \,\delta_+((p_k - q)^2 - m_k^2) \;,
\eeq
gives the corresponding imaginary part (see, e.g., Ref.~\cite{be90}). 
Note that in the massive case
the imaginary part is more involved than in the massless case, where it is obtained
from the real part by the simple analytic continuation
$\ln(s_{jk}) \to \ln (-s_{jk} - \i 0) = \ln |s_{jk}| - \i\pi\Theta(s_{jk})$
of its overall factor $({\mu^2}/{s_{jk}})^{\ep}$ in Eq.~(\ref{iee}).

According to our definition of the insertion operator
$\bom I$ in  Eq.~(\ref{ff1loop}), the finite one-loop contribution 
$\cm_m^{(1)\, {\rm fin}}$ still depends on the dimensional-regularization scale
$\mu$. 
%
This dependence is nonetheless simple, because it is embodied in a contribution
proportional to $\ln \mu$. The coefficient of this single-logarithmic 
contribution is given by
\beq
\label{mudep}
\mu^2 \frac{d}{d\mu^2} \,\cm_m^{(1)\, {\rm fin}}(\mu^2;\{p_i,m_i\}) = 
\left( q\,\frac{\beta_0}{2} - 
\sum_{j=1}^{m} \left[ \gamma_j -\frac{2}{3} T_R N_F \delta_{jg} \right]
\right) \cm_m^{(0)}(\{p_i,m_i\}) \;,
\eeq 
Note that $\mu$ should not be confused with the renormalization scale.
In particular, if renormalized masses (and related Yukawa couplings) do
not correspond to the pole-mass definition, but they are \msbar-scheme
running masses at the renormalization scale $\mu_R$, the finite
contribution $\cm_m^{(1)\, {\rm fin}}$ can explicitly contain
terms of the type $m_i \ln (m_i/\mu_R)$.

\subsection{SUSY QCD amplitudes}
\label{sing1loopSUSYQCD}

All the results on QCD amplitudes in Sect.~\ref{sing1loopQCD} can be
extended to SUSY QCD processes. A preliminary discussion on the RS and
on the small-mass limit is nonetheless necessary.

For a supersymmetric theory CDR (in contrast to DR) is not a consistent
RS, because the mismatch between the $d-2=2(1-\ep)$ transverse degrees
of freedom of the gluons and the 2 transverse degrees of freedom of the 
gluinos violates supersymmetric Ward identities. In particular, in the
case of on-shell one-loop amplitudes, this leads to violation of the
tree-level identity $\gs = {\hat \gs}$, between the gluon (gauge)
coupling $\gs$ and the $q \tq \tg$-Yukawa coupling ${\hat \gs}$.
Nevertheless, CDR can also be used in SUSY calculations, because
supersymmetry can be restored by introducing a proper counterterm
\cite{MPrestore,BHSZsusyQCD}. More precisely, we can still use the
notation of Eq.~(\ref{loopex}) (where only $\gs$ appears) in any RS,
provided we implement the following RS-dependent relation between the
two couplings at one-loop order:
\beq
{\hat \gs}^{\RS} = \gs \left[ 1 + \left( \frac{\gs}{4\pi} \right)^2 
{\hat \gamma}^{\RS} \right] \;\;,
\eeq
where 
\beq
{\hat \gamma}^{\DR} =0 \;, \quad \quad 
{\hat \gamma}^{\CDR} = \frac{2}{3} \,C_A - \frac{1}{2} \, C_F \;.
\eeq

As discussed in detail in Sect.~\ref{sing1loopQCD}, our factorization formula
for one-loop QCD amplitudes smoothly interpolates between the cases of massless
and massive quarks. Analogous results, with smooth behaviour with respect to
sparticle masses, can be obtained for SUSY QCD. However, in this section we
are not going to present such general results because of the following reasons.
On one hand, the case of exactly massless SUSY partons has no practical
interest. One the other hand, owing to the proliferation of masses in
the SUSY particle spectrum, the control of logarithmically-enhanced
terms produced by the smallness of the mass of some SUSY partons would
require a quite involved presentation and detailed discussions of the
various possible cases related to different scenarios of mass hierarchies.

For the sake of simplicity, we thus limit ourselves to considering
the case of finite (non-vanishing) squark and gluino masses, $m_\tq$
and $m_\tg$. The factorization formula given below includes {\em all}
the divergent $\ep$-poles that appear in this case, and {\em some}
related logarithmic terms.  

The one-loop amplitudes for SUSY QCD processes have the same
singularity structure as given by Eqs.~(\ref{ff1loop}) and (\ref{iee}).
The coefficient $\beta_0$ in Eq.~(\ref{iee}) should simply be replaced
by $\beta_0^{\rm SUSY}$, the first coefficient of the SUSY QCD beta
function:
\beq
\label{beta0SUSY}
\beta_0^{\rm SUSY} =
\frac{11}{3}\,C_A - \frac{4}{3}\,T_R ( N_f + N_F )
- \frac{2}{3}\,T_R N_S - \frac{2}{3}\,C_A\:,
\eeq
where $N_S$ is the number of chiral squark pairs $\tq=(\tq_R,\tq_L)$
($N_S= N_f + N_F$ in the fully supersymmetric theory).  Moreover, the
sums over the flavour indices $j,k$ in Eq.~(\ref{iee}) run over all
parton species $(j,k=q,{\bar q},g,\tq,\tg)$. Correspondingly,
we have to introduce the flavour functions $\Gamma_j$ of the
external massive gluinos and squarks:
\beq
\label{cgammaspm}
\Gamma_j(\mu,m_j;\ep)= 
{\bom T}_j^2 \left( \frac{1}{\ep} - \ln\frac{m_j^2}{\mu^2} -2 \right) +
\gamma_j \,\ln\frac{m_j^2}{\mu^2} \;, \quad j=\tg,\tq \;,
\eeq
where
\beq
\label{cgammas}
\gamma_{\tg} = \frac{3}{2} \, C_A \;,
\qquad\qquad\quad\;\;
\gamma_\tq = 2 \,C_F \:.
\eeq

As discussed above, the extension of Eq.~(\ref{iee}) to SUSY QCD by
using Eqs.~(\ref{beta0SUSY})--(\ref{cgammas}) does not include in the
insertion operator ${\bom I}$ all the terms that are
logarithmically-enhanced in the limit of small squark and gluino
masses. Note however, that the finite contribution
$\cm_m^{(1)\, {\rm fin}}(\mu^2;\{p_i,m_i\})$ of the one-loop SUSY QCD
amplitude still fulfils the property in Eq.~(\ref{masslesslimit}),
provided the massless limit $m_i \to 0$ is restricted to quarks and/or
antiquarks $(i=q,{\bar q})$.  

The $\mu$-dependence of $\cm_m^{(1)\, {\rm fin}}(\mu^2;\{p_i,m_i\})$
is still given by Eq.~(\ref{mudep}), with the replacement $\beta_0 \to
\beta_0^{\rm SUSY}$ and the inclusions of the flavour coefficients 
$\gamma_{j=\tg,\tq}$ in Eq.~(\ref{cgammas}).

\section{Relation of the singular terms to the Altarelli-Parisi
splitting functions}
\label{APrelation}

In the case of massless QCD, it is known \cite{GG, KSTsing, CSdipole}
that  the flavour functions $\Gamma_j^{\RS}$ of Eq.~(\ref{iee}) are
related to the \AP\ splitting functions. In this section we first
recall this correspondence and then we sketch how it can be extended to
the massive case.  

Let us consider, for instance, the collinear splitting of a massless quark.
The corresponding \AP\ splitting function ${\hat P}^{\RS}_{qq}(z;\ep)$
in the dimensionally-regularized theory can be written as
\beq
\label{pqq0}
{\hat P}_{qq}^{\RS}(z;\ep) =C_F\,
\left[ \frac{1+z^2}{1-z} + \frac{h_g^{\RS}(\ep) -2}{2} (1-z) \right]
\equiv \frac{2 C_F}{1-z} + {\hat P}_{qq}^{({\rm reg})\RS}(z;\ep) \:,
\eeq
where $h_g^\RS(\ep)$ is the number of gluon helicity states in the
corresponding RS (see Eq.~(\ref{hgscheme})). On the right-hand side we
have isolated the contribution that becomes singular in the soft region
$z \to 1$ from the remaining regular part ${\hat P}_{qq}^{({\rm reg})\RS}$. 

The soft contribution%
\footnote{Analogous soft terms, which become singular when $z \to 0$,
appear in the \AP\ functions of other splitting processes (see the
equations in the Appendix).} 
has to be
combined with the soft terms produced
by non-collinear (large-angle) gluon radiation (see Eq.~(\ref{3pfun})). 
The terms that are singular both in the soft and collinear regimes lead
to the double-pole singularity of the colour-correlation function in
Eq.~(\ref{cVmasslessj}). In the massive case kinematics is different
from the massless case, and the contribution of these soft terms is
embodied in
the functions $\cV^{(\rm cc)}_{jk}(s_{jk}; m_j,m_k;\ep)$,
which control the entire colour-correlation part of Eq.~(\ref{iee}).

The term ${\hat P}_{qq}^{({\rm reg})\RS}$ can only lead to collinear
singularities (single poles and 
constants) and it is related to the function
$\Gamma_q^{\RS}$
in Eq.~(\ref{cgammaq0}), 
including its RS dependence \cite{CSTrsdep}:
\beeq
\label{gammaint}
\gamma_q &=& - \int_0^1\!\d z \;{\hat P}_{qq}^{({\rm reg})}(z;\ep=0) \:, \\
\tilde{\gamma}_q^{\RS} - \tilde{\gamma}_q^{\CDR} &=& \int_0^1\!\d z
\;\frac{1}{\ep} 
\left[ 
{\hat P}_{qq}^{({\rm reg})\RS}(z;\ep) - {\hat P}_{qq}^{({\rm reg})\CDR}(z;\ep)
\right] \:.
\eeeq
Analogous equations relate the other flavour coefficients $\gamma_j$, 
$\tilde{\gamma}_j^{\RS}$ in Eqs.~(\ref{cgamma}), (\ref{cgammatilde}), 
(\ref{cgammas}) to the corresponding terms
${\hat P}_{jk}^{({\rm reg}) \RS}$ of the \AP\ splitting functions.

The relation between the functions $\Gamma_j^{\RS}$ and the splitting
functions can be extended from the massless to the massive case, 
provided we properly take into account the corresponding 
{\em dynamics} and kinematics differences.

The dynamics of the splitting processes of massive partons can be
described by generalizing the collinear limit to the
{\em quasi-collinear limit}.  Let us consider a generic tree-level
amplitude $\cm_{m+1}^{(0)}(\{p_i,m_i\})$ with $m+1$ external partons.
The limit when an internal parent parton (labelled by $(jk)$) decays
quasi-collinearly in two external partons $j$ and $k$ is defined by
\beq
\label{qclim}
p_j^\mu \to z p^\mu \;\;, \quad\quad p_k^\mu \to (1-z) p^\mu \;\;, 
\quad\quad p^2 = m^2_{(jk)} \;\;,
\eeq
with the constraint
\beq
p_j\cdot p_k, \; m_j, \; m_k, \; m_{(jk)} \to 0 
\quad\quad {\rm at \; fixed \; ratios} \quad\quad
\frac{m_j^2}{p_j\cdot p_k},\;\frac{m_k^2}{p_j\cdot p_k}, \;
\frac{m_{(jk)}^2}{p_j\cdot p_k} \:.
\eeq
The quasi-collinear limit obviously differs from the collinear limit
because the splitting partons are massive. However, the key difference
between the two limits is given by the constraint that the on-shell
masses squared have to be kept of the same order as the invariant mass
$(p_j+p_k)^2$, while the latter become small.  

It can be shown that in the quasi-collinear limit the tree-level
squared amplitude fulfils a factorization formula similar to the
analogous formula for the collinear limit. We have
\beq
\label{qcfac}
|\cm_{m+1}^{(0)}|^2 \;\sim\; |\cm_{m}^{(0)}|^2
\,\frac{2\mu^{2\ep}\gs^2}{(p_j+p_k)^2 - m_{(jk)}^2}
\;{\hat P}_{(jk),j}^\RS(z;\ep; \{ \mu_{l}^2 \}) \:,
\eeq
where the $m$-parton matrix element on the right-hand side 
is obtained from $\cm_{m+1}^{(0)}$ by replacing the partons $j$ and $k$
with the single parent parton of momentum $p$.
The function ${\hat P}_{(jk),j}^\RS(z;\ep;\{ \mu_{l}^2 \})$ generalizes
the customary $d$-dimensional \AP\ splitting function to the case of
quasi-collinear splitting. It has the usual dependence on $\ep$ and on
the longitutinal-momentum fraction $z$, plus an additional dependence
on the parton masses indicated by the variables $\{ \mu_{l} \}$:
\beq
\label{massdep}
\mu_{l}^2 = \frac{m_l^2}{(p_j+p_k)^2-m_{(jk)}^2} \;\;,
\eeq
where (in general) $m_{l}^2$
stands for any quadratic combination of the 
masses of the partons involved in the splitting process 
$(m_{l}^2 = m_j^2,m_k^2,m_{(jk)}^2,m_jm_k, \dots$ and so forth).

For instance, in the case of the quasi-collinear splitting $q \to q + g$
of a massive quark $(p_q^2=m_q^2, \; p_g^2=0)$, the analogue of the massless
splitting function in Eq.~(\ref{pqq0}) is 
(see Eq.~(\ref{avhpqq}) and Refs.~\cite{Keller:1999tf,di00}):
\beq
\label{pqqm}
{\hat P}_{qq}^{\RS}\left(z;\ep;\frac{m_q^2}{2p_q p_g}\right) = 
\frac{2 C_F}{1-z} + \left[ {\hat P}_{qq}^{({\rm reg})\RS}(z;\ep)
 - C_F \frac{m_q^2}{p_q p_g} \right] \:.
\eeq
The `regular' part of the splitting function is now given by the
contribution in the square bracket. It contains a term that explicitly
depends on $m_q$, in addition to the regular part that appears in the
massless case (see Eq.~(\ref{pqq0})).

Comparing the regular parts of the splitting functions in
Eqs.~(\ref{pqq0}) and (\ref{pqqm}), we can understand the difference
between the the massless and massive functions $\Gamma_q$ in
Eqs.~(\ref{cgammaq0}) and (\ref{cgammaqm}). This difference amounts to
the correspondence
\beq
\label{singleepspole}
\frac{1}{\ep} \left( \gamma_q - \ep {\tilde \gamma}^{\RS}_q \right)
\longleftrightarrow \gamma_q
\ln\frac{m_q^2}{\mu^2} + C_F \frac{1}{\ep} \left( 1
- \ep \ln \frac{m_q^2}{\mu^2} - 2 \ep \right) \:.
\eeq
In both the massless and massive cases, $\gamma_q$ is given by
Eq.~(\ref{gammaint}).  The coefficient in front of $\gamma_q$ is
obtained by performing the integration of the propagator factor
$1/p_qp_g$ (see Eq.~(\ref{qcfac})) over the relative angle
$\theta_{qg}$ between the quark and the gluon. In the massless case
this integration is singular when $\theta_{qg} \to 0$ and it leads to
the single pole $1/\ep$ on the left-hand side. In the massive case, the
integration is kinematically regularized at a cutoff value
$\theta_{qg} \gtap m_q/E_q$ and it leads to the logarithmic behaviour
of the first term on the right-hand side.  The contribution inside the
round bracket on the right-hand side is instead produced by the term of
Eq.~(\ref{pqqm}) that explicitly depends on $m_q$. Here, the angular
integration, which is dominated by the region $\theta_{qg} \ltap m_q/E_q$,
produces a constant term,
 while the integration over the gluon energy
$E_g \sim (1-z) E_q$ is divergent in the soft region $z \to 1$ and it
leads to the single pole $1/\ep$ in front of the round bracket.

The generalized \AP\ functions for the quasi-collinear limit of the
other splitting processes in QCD and SUSY QCD are listed in the
Appendix.  The expressions of the flavour functions in
Eq.~(\ref{cgammag}) and Eq.~(\ref{cgammaspm}) are related to the
splitting functions in Eqs.~(\ref{avhpgq}), (\ref{avhpgg}) and 
Eqs.~(\ref{avhptgtg}), (\ref{avhptqtq}), respectively.

\section{Summary}
\label{sum}

In this paper we have discussed the singular behaviour of on-shell QCD
and SUSY QCD amplitudes at one-loop order in the presence of massive
particles. The complete structure of the ultraviolet and infrared
singularities is described by the colour-space factorization formula
given in Sect.~\ref{sing1loopmassive}. The factorization formula is
universal, i.e.~valid for any amplitude, and the singular factors only
depend on the flavours and momenta of the coloured external legs. 
Moreover, the factorization formula is given in such a form that the
corresponding formula for massless QCD partons is smoothly recovered by
simply letting the masses approach to zero. 

Our factorization formula can be useful both to check explicit
evaluations of one-loop amplitudes and to organize their calculations
in terms of divergent parts and finite remainders.
Furthermore, in the asymptotic regime where the parton masses are much
smaller than any of the relevant kinematic invariants, the formula can
also be used to directly obtain (apart from vanishing corrections when
the masses tend to zero) the one-loop massive amplitude from the
corresponding massless amplitude, without explicitly computing the
former. 
In the general
context of NLO calculations of jet observables, our one-loop results
are useful for setting up the integration of tree-level amplitudes in
such a way as to construct process-independent techniques for infrared
cancellations.

\noindent {\bf Acknowledgements}. 
We thank Mike Seymour for collaboration at an early stage of this work.
Z.T. is grateful to the
CERN Theory Division for the hospitality while this work was performed.
This work was supported in part by the EU Fourth Framework Programme
`Training and Mobility of Researchers', Network `Quantum Chromodynamics
and the Deep Structure of Elementary Particles', contract
FMRX-CT98-0194 (DG 12 - MIHT) as well as by the Hungarian Scientific
Research Fund grant OTKA T-025482.

\section*{Appendix: Quasi-collinear dynamics and generalized \AP\ 
splitting functions}
\label{APfunctions}

In this appendix we list \AP splitting
functions in QCD and SUSY QCD, supplemented by the mass terms that are
relevant in the quasi-collinear limit discussed in
Sect.~\ref{APrelation} 
(see Eqs.~(\ref{qclim})-(\ref{massdep})). In the 
expressions given below all the mass terms can be parametrized by the variable
$\mu^2_{jk}= (m_j^2+m_k^2)/[(p_j+p_k)^2-m_{(jk)}^2]$. We include the dependence
on the RS, which is parametrized by the corresponding number $h_g^\RS(\ep)$  
of gluon polarizations,
\beq
\label{hgscheme}
h_g^{\rm CDR} = 
d - 2 = 2 - 2\ep\:,\qquad h_g^{\rm DR} = 2\:.
\eeq
Note that our expressions refer to \AP\ splitting functions after 
average over the azimuthal angle identified by the collinear direction.
The azimuthally averaged functions are relevant to discuss the RS dependence
\cite{CSTrsdep} and, in general, they do not coincide with the splitting 
functions averaged over the polarizations of the parent parton. 
Since we are considering the splitting process $(jk) \to j \,k$,
the splitting functions fulfil the obvious symmetry relation
${\hat P}^\RS_{(jk)k}(z;\ep;\mu_{kj}^2)=
{\hat P}^\RS_{(jk)j}(1-z;\ep;\mu_{jk}^2)$.

{\bf Quarks and gluons:}
\beeq
&&
\label{avhpqq}
\average{ {\hat P}^\RS_{qq}(z;\ep;\mu_{qg}^2) } = C_F\,
\left[ \frac{2z}{1-z} + \frac12\,h_g^\RS (1-z)
     - 2\mu_{qg}^2\right] \:,
\\ &&
\label{avhpgq}
\average{ {\hat P}^\RS_{gq}(z;\ep;\mu_{q{\bar q}}^2) } = T_R\,
\left[ 1 -
\frac{2}{d-2}\left(2 z(1-z) - \mu_{q{\bar q}}^2\right)
\right] \:,
\\ &&
\label{avhpgg}
\average{ {\hat P}^\RS_{gg}(z;\ep) } = 2C_A\,
\left[ \frac{z}{1-z} + \frac{1-z}{z} + \frac{h_g^\RS}{d - 2} z(1-z) \right] \:.
\eeeq

{\bf Gluinos and gluons:}
\beeq
&&
\label{avhptgtg}
\average{ {\hat P}^\RS_{\tg\tg}(z;\ep;\mu_{\tg g}^2) } = C_A\,
\left[ \frac{2z}{1-z} + \frac12\,h_g^\RS (1-z)
     - 2\mu_{\tg g}^2\right] \:,
\\ &&
\label{avhptgq}
\average{ {\hat P}^\RS_{g\tg}(z;\ep;\mu_{\tg\tg}^2) } = C_A\,
\left[ 1 - \frac{2}{d-2}\left(2 z(1-z) - \mu_{\tg\tg}^2\right) \right] \:.
\eeeq

{\bf Squarks and gluons:}
\beeq
&&
\label{avhptqtq}
\average{ {\hat P}^\RS_{\tq\tq}(z;\ep;\mu_{\tq g}^2) } = C_F\,
\left[ \frac{2z}{1-z} - 2\mu_{\tq g}^2\right] \:,
\\ &&
\label{avhpgtq}
\average{ {\hat P}^\RS_{g\tq}(z;\ep;\mu_{\tq{\bar{\tq}}}^2) } = T_R\,
\frac{1}{d-2}\left[2z(1-z) - \mu_{\tq{\bar{\tq}}}^2 \right] \:.
\eeeq
Equations~(\ref{avhptqtq})--(\ref{avhpgtq}) are valid for the superpartners of
both the left- and the right-chirality quarks.

Neglecting the mass terms, which are proportional to $\mu_{jk}^2$, the
splitting functions coincide with those reported in 
Ref.~\cite{BHSZsusyQCD} for the CDR scheme. We do not consider the
splitting functions produced by the Yukawa coupling $q\tg\tq$,
because, as long as the gluino and squark masses are finite, they do
not produce any singular terms when $\ep \to 0$.

\def\ac#1#2#3{Acta Phys.\ Polon.\ #1 (#3) #2}
\def\ap#1#2#3{Ann.\ Phys.\ (NY) #1 (#3) #2}
\def\ar#1#2#3{Annu.\ Rev.\ Nucl.\ Part.\ Sci.\ #1 (#3) #2}
\def\cpc#1#2#3{Computer Phys.\ Comm.\ #1 (#3) #2}
\def\ib#1#2#3{ibid.\ #1 (#3) #2}
\def\np#1#2#3{Nucl.\ Phys.\ B#1 (#3) #2}
\def\pl#1#2#3{Phys.\ Lett.\ #1B (#3) #2}
\def\plb#1#2#3{Phys.\ Lett.\ B#1 (#3) #2}
\def\pr#1#2#3{Phys.\ Rev.\ D #1 (#3) #2}
\def\prep#1#2#3{Phys.\ Rep.\ #1 (#3) #2}
\def\prl#1#2#3{Phys.\ Rev.\ Lett.\ #1 (#3) #2}
\def\rmp#1#2#3{Rev.\ Mod.\ Phys.\ #1 (#3) #2}
\def\sj#1#2#3{Sov.\ J.\ Nucl.\ Phys.\ #1 (#3) #2}
\def\zp#1#2#3{Z.\ Phys.\ C#1 (#3) #2}

\end{document}